# Ising domain wall networks from intertwined charge density waves in single-layer TiSe$_2$

Wen Wan[1], Maria N. Gastiasoro[1], Daniel Muñoz-Segovia[1], Paul Dreher[1],

Miguel M. Ugeda*,[1,2,3] and Fernando de Juan*,[1,3]

[1]Donostia International Physics Center, Paseo Manuel de Lardizábal 4, 20018 Donostia-San Sebastián, Spain

[2]Centro de Física de Materiales, Paseo Manuel de Lardizábal 5, 20018 San Sebastián, Spain.

[3]Ikerbasque, Basque Foundation for Science, Plaza Euskadi 5, 48009 Bilbao, Spain

*Corresponding authors: mmugeda@dipc.org, fernando.dejuan@dipc.org

*When the period of an incommensurate charge density wave (ICDW) approaches an integer multiple of a lattice vector, the energy gain obtained from locking the period to the lattice can lead to a fascinating transition into a commensurate state. This transition actually occurs through an intermediate near-commensurate (NC) phase, with locally commensurate regions separated by an ordered array of phase slips of a complex CDW order parameter. TiSe$_2$ is a paradigmatic CDW system where incommensuration is believed to be induced by carrier doping, yet its putative NC state has never been imaged or its nature established. Here we report the observation of a striking NC state in ultraclean, slightly doped monolayers of TiSe$_2$, displaying an intricate network of coherent, unidirectional CDW domain walls over hundreds of nanometers. Detailed analysis reveals these are not phase slips of a complex CDW, but rather sign-changing Ising-type domain walls of two coupled real CDWs of previously known symmetry, consistent with the period doubling nature of the parent commensurate state. In addition, we observe an unexpected nematic modulation at the original lattice Bragg peaks which couples to the CDW order parameters. A Ginzburg-Landau analysis naturally explains the couplings and relative modulations of all order parameters, unveiling TiSe$_2$ as a rare example of an NC-CDW of two intertwined real modulations and emergent nematicity.*

Understanding how ICDW states lock to the crystal lattice in an incommensurate to commensurate (I-C) transition is a fascinating problem with profound implications on how other ordered states such as superconductivity emerge in their presence[1]. Due to the energetics of lattice locking, the I-C transition typically proceeds via an intermediate near-commensurate (NC) phase characterized by locally commensurate regions separated by domain walls (see phase diagram in Fig. 1b). The nature of these domain walls, and of the I-C transition itself, depends on whether the CDW order parameter is real or complex, which is determined by the commensurate CDW (CCDW) period[2]. In a one-dimensional CCDW with wavevector $Q = 2\pi/Na$, where $a$ is the lattice constant, lattice translations for $N > 2$ are represented by a phase factor $e^{i2\pi/N}$ and the order parameter is therefore complex. In this case, the NC state has approximately constant amplitude but shows regions of constant phase separated by phase slips known as discommensurations[3], akin to domain walls of easy-plane ferromagnets. If the CCDW has period $N = 2$ however, translations are represented by a sign change ($e^{i2\pi/N}$ = -1) and the order parameter is real. In this case, only real sign-changing domain walls where the order parameter goes through zero are possible, akin to domain walls in Ising ferromagnets (Supplementary Fig. S1).

TiSe$_2$ is a paradigmatic example of one such $N = 2$ triple-Q CDW, which can be driven through a C-I transition by electronic doping. The bulk CDW wavevector is $Q = 2\pi \left(\frac{1}{2a}, 0, \frac{1}{2a}\right)$ and Cu doping[4] or applying pressure[5] leads to an out-of-plane incommensurate CDW state which becomes superconducting at $T_C$ ~2-4 K (ref. 6). In-plane incommensuration is not resolved in X-rays, but STM studies in Cu- and Pt-doped samples do observe in-plane CDW domain walls[7–11], with average separation of 10 nm, and a disordered spatial distribution likely dominated by CDW pinning by the Cu dopants. Indirect transport signatures of incommensuration appear in gated thin films as well[12], but to prove the existence of an intrinsically modulated, coherent in-plane NC-CDW state, imaging of a homogeneously doped sample without disorder would be required. To disentangle the more complex out-of-plane incommensuration, this should ideally be done in the single-layer limit, where the $N = 2$ CDW remains essentially unchanged from the bulk[13–19].

In this work, we demonstrate that a doped monolayer of 1T-TiSe$_2$ exhibits an intrinsic in-plane NC-CDW state with wavelength of 20 nm, which remains coherent over hundreds of nanometers. We uncover the spatial structure of the NC state via high-resolution STM measurements combined with standard phase-locking techniques, and show it is made of a train of intertwined Ising domain walls of two independent order parameters with opposite mirror parity (shown schematically in Fig. 1c). These domain walls concatenate four CDW domains (locally commensurate CDW ground states) out of the eight available ones (A through D' in Fig. 1d) in a periodic arrangement. Our analysis also reveals the existence of a strong nematic modulation of the lattice coupled to the domain wall train. This complex NC state is naturally accounted for by a Ginzburg-Landau theory, including the order and number of the domains and their coupling. Our work clearly shows how TiSe$_2$ proceeds through the doping-driven C-I phase transition via an NC state of Ising domains.

**Near commensurate CDW state**

Our experiments were carried out on a nearly full monolayer of 1T-TiSe$_2$ grown on epitaxial bilayer graphene (BLG) on 6H-SiC(0001), as sketched in Fig. 2a. The large-scale morphology of our TiSe$_2$ monolayers is characterized by a remarkable uniformity with single-crystal domains of several hundreds of nanometers in size (Fig. 2b) with a density of point defects below $1\cdot 10^{12}$ cm$^{-2}$ (see Methods and Supplementary Fig. 3). A typical STM d$I$/d$V$ spectrum, showing low-lying electronic structure of single-layer TiSe$_2$, is shown in Fig. 2c (for a larger energy range, see Supplementary Section 3). Three step-like features for occupied states are observed at $V_s$ = -215 ± 5 mV (labeled $V_1$), $V_s$ = -365 ± 5 mV (labeled $V_2$), and a subtler step with opposite orientation (see inset) labeled $C_1$ at $V_s$ = -65 ± 5 mV. These features match the band structure of monolayer TiSe$_2$ in the CDW phase (Fig. 2d adapted from ref. [13]), which shows an electron band of Ti-3$d$ character (back-folded from M) and two spin-orbit split Se-4$p$ hole bands ($E_{SOC}$ = 150 ± 10 meV) (Fig. 2d) separated by the CDW gap ($E_G$ = 150 ± 10 meV)[20–25]. This identification is also supported by previous angle-resolved photoemission spectroscopy (ARPES) measurements[13]. Importantly, both the ARPES and STS measurements reveal a small electron doping because the $d$-electron band (upper band in Fig. 2d) is partially filled, likely due to charge transfer from the substrate.

Our atomically resolved STM images of the CDW state show the expected 2x2 order in small scales (Fig. 2e), where the CDW unit cell is doubled in both lattice directions (white hexagon) compared to the crystal unit cell (green hexagon). The corresponding FFT in Fig. 2f shows clear superlattice Bragg peaks at both the $\bar{Q}_n = \bar{M}_n = \bar{G}_n/2$ point ($n$ = 1,2,3) and the higher harmonic $\bar{M}'_n = \bar{M}_n + \bar{G}_{n-1}$ (yellow and orange circles in Fig. 2g, respectively) which are the signature of 2x2 order. While this 2x2 modulation seen in STM has traditionally been identified[7,26–29] as resulting from the 2x2 lattice distortion observed in neutron scattering[30], we now show the STM modulation actually reflects the existence of two CDW order parameters of different symmetry.

The primary CDW order parameter of monolayer TiSe$_2$ (point group symmetry D$_{3d}$), is known to be a transverse phonon of symmetry $M_1^-$ (irrep A$_u$ of the little group C$_{2h}$ at $M$), which is *odd* under both inversion and the mirror symmetry $\sigma_v$ parallel to the corresponding $\bar{M}_n$ vector. A depiction of the triple-Q $M_1^-$ phonon distortion is shown in Fig. 1c, with in-plane black arrows representing the displacements of the top Se layer. A secondary CDW order parameter of $M_1^+$ symmetry, representing a modulation where one out of four Se atoms rises upwards, is symmetry allowed and condenses simultaneously at the CDW transition[31]. The corresponding $M_1^+$ distortion, *even* under both inversion and mirror $\sigma_v$, is also depicted in Fig. 1c (out-of-plane black arrows). Note in the monolayer, an $M_1^-$ distortion generically makes the ground state chiral[32], unlike in the bulk, where $L_1^-$ always preserves an inversion center between layers, because chirality alternates from one layer to the next. In our monolayer samples, the $M_1^-$ and $M_1^+$ modulations can be distinguished in STM by their mirror parity as follows. The charge density in the CDW state is approximately given by $\rho(x) = 2\Re[\sum_n A_n^G e^{i\bar{G}_n \bar{x}} + A_n^M e^{i\bar{M}_n \bar{x}} + A_n^{M'} e^{i\bar{M}'_n \bar{x}}]$, where $A_n^q$ are complex numbers at the $\bar{q} = \bar{G}_n, \bar{M}_n$ and $\bar{M}'_n$ FFT peaks. Note because lattice translations act as real numbers $e^{iq_i \bar{a}_j} = \pm 1$ for any $q_i$, the real and imaginary parts of $A_n^q$ encode independent 2x2 modulations, which may even have different symmetry, as we show in the following. Since $\bar{M}_n (\bar{M}'_n)$ is parallel (perpendicular) to $\sigma_v$, the modulation $e^{i\bar{M}_n \bar{x}}$ is invariant under $\sigma_v$, while for $M'$ $e^{i\bar{M}'_n \bar{x}} \to e^{-i\bar{M}'_n \bar{x}}$. This implies that $\Re A_n^M$, $\Im A_n^M$ and $\Re A_n^{M'}$ all represent mirror even 2x2 modulations, while only $\Im A_n^{M'}$ represents a mirror odd 2x2 modulation. Hence, the primary CDW order parameter, which we call $\Delta_n$, can only be $\Delta_n = \Im A_n^{M'}$, which is sensitive to the small in-plane modulations of the Se atoms of symmetry $M_1^-$. The other three mirror even signals ($\Re A_n^M$, $\Im A_n^M$ and $\Re A_n^{M'}$) can be rearranged into three $M_1^+$ order parameters $\phi_n^\alpha$ showing the 1-in-4 pattern localized at the upper Se ($\phi_n^{SeUp}$), Ti ($\phi_n^{Ti}$) and lower Se ($\phi_n^{SeDo}$) sites, respectively (Supplementary Section 4).

To extract all these order parameters, we implemented the Lawler-Fujita (LF) algorithm to produce corrected images in a perfect registry with the lattice, which allows to extract the complex numbers $A_n^M$ and $A_n^{M'}$ with well-defined phases (Methods). Fig. 2g shows a one-dimensional cut of the extracted $M_1^+$ secondary order parameters $\phi^{SeUp}$, $\phi^{Ti}$, $\phi^{SeDo}$, showing the dominance of $\phi^{SeUp}$ as this layer is the closest to the surface and, therefore, to the STM tip. Because of this, from now on we only consider $\phi_n \equiv \phi_n^{SeUp}$ as the secondary CDW order parameter. The extracted primary and secondary order parameters, $\Delta_n$ and $\phi_n$, along the same one-dimensional cut are also shown in Fig. 2g, revealing finite $n = 1,2,3$ components of the triple-Q CDW state.

As a check, in Fig. 2h we reconstruct an approximate $\rho(\bar{x})$ using the extracted order parameters of Fig. 2g, obtained by averaging the spatially dependent ones (Supplementary Section 4). In remarkable agreement with the STM image (Fig. 2e), Fig. 2h shows the same weakly rotated propeller-like shape and the 1-in-4 bright feature of the Se atoms. The rotation of this propeller reflects the existence of $\Delta_n$, with its chirality (clockwise vs counter-clockwise) given by the sign of $\Delta_1 \Delta_2 \Delta_3$, while the 1-in-4 bright upper Se atom originates from $\phi_n$. The four choices in the CDW unit cell for the propeller center, determined by the signs of $\Delta_n$ and $\phi_n$, make eight possible CCDW ground states, which we label A,B,C,D for one chirality, and A',B',C',D' for the other, represented schematically in Fig.1d. The local patch in Fig. 2e therefore corresponds to state A. Note that to determine without ambiguity the chirality of $\Delta_n(x)$ experimentally, all three sublattices (Se-up, Ti, Se-down) have to be resolved, as this sets the coordinate system and the reciprocal lattice vectors. A faithful extraction of the primary $\Delta_n$ and secondary $\phi_n$ order parameters, done for the first time here, is crucial to understand the symmetry and domain structure of the CDW ground state, as well as any potential incommensurate state.

The main experimental result of our work is now unveiled by our high-resolution topography images taken over very large scales, which show a highly inhomogeneous 2x2 CDW displaying a striking long-range ordering of domains characteristic of a near-commensurate phase. Figure 3a shows an atomically resolved STM image of 153 nm x 153 nm in size (2048 px. × 2048 px., 0.7 Å/px.), where long striped regions dominate the topography, forming an intricate 2D modulated pattern. The corresponding FFT of this STM image, shown in Fig.3b, reveals characteristic CDW peaks at $M_n$ and $M'_n$ as observed for a single domain (Fig. 2f). While correcting the phase drift with the LF algorithm in such a large, highly modulated image is not possible, the Fourier-filtered amplitude (insensitive to phase drift) of the $M$ peaks (Fig. 3b), reveals that the long-wavelength topography modulation indeed comes from a modulation of the CDW, and shows the wavelength appears halved compared to the topography. We systematically observed these patterns in tens of different regions in several TiSe$_2$/BLG samples (Supplementary Fig. S7), enabling us to extract an averaged periodicity of $\lambda = 20 \pm 2$ nm for the modulation induced perpendicular to the domain walls. Remarkably, the incommensuration is single-q in character, i.e. the modulation produces 1D stripes, and is not a 2D network as believed previously[12,33]. This CDW is insensitive to the application of out-of-plane magnetic fields (Supplementary Fig. S8).

The one-dimensional trains of domains in Fig. 3b form 'super-domains' that can be distinguished by the modulation vector $\bar{q}_{NC}$, roughly parallel to one of the CDW wave-vectors $\bar{Q}_i$. When the super-domains meet, domain walls twist and cross, possibly being pinned by a lattice defect. Figure 3d shows a zoom of the three types of super-domains. These images also reveal a finer feature of domain walls: there are always two classes of them, wide and narrow, alternating within the super-domain train. The behavior of the CDW in the single-layer limit contrasts with the case of pristine bulk TiSe$_2$, where the CDW shows a commensurate, single-domain phase[28,34]. The multidomain structure of CDW order in the single layer cannot be attributed to disorder given the low density of defects of our films (Supplementary Fig.S3), and rather reveals an intrinsic near-commensurate state induced by substrate charge transfer. These patterns can be still seen at 77 K but are not present above $T_{CDW} \approx 230K$, as seen in our STM images at room temperature (Supplementary Fig. S9), which also ensures they are not extrinsic and originate from the incommensuration of the CDW.

**Intertwined CDW order parameters**

To analyze the order parameter structure of the NC state we selected a STM image with a single super-domain (Fig. 4a). The extracted order parameters are shown in Fig. 4b, which reveals another striking feature: the modulation of the order parameter differs in the different $\Delta_n$ components. Taking super-domains where $\bar{q}_{NC} \propto \bar{G}_2$ as an example, to a very good approximation we have $\Delta_2 \propto sin(2\bar{q}_{NC}\bar{x})$ while $\Delta_1 \propto sin(\bar{q}_{NC}\bar{x}) + c \cdot sin(3\bar{q}_{NC}\bar{x})$ and $\Delta_3 \propto cos(\bar{q}_{NC}\bar{x}) - c \cdot cos(3\bar{q}_{NC}\bar{x})$ with $|c|$

< 1. The secondary order parameter components follow a very similar pattern with the opposite sign, $\phi_n = -\Delta_n$. This leads to sign alternation of domains which produces the pattern "ABCD" illustrated in Fig. 4d, where the four CCDW states with counter-clockwise chirality $\Delta_1\Delta_2\Delta_3 < 0$, (left column in Fig. 1d) are interspersed within the super-domain. A zoom around a domain wall in Fig. 4c shows how a given pattern of sign changes of the order parameters shifts the center of the CDW by one atomic unit cell: in this case, the transition from state B to C corresponds to a translation by lattice vector $\bar{a}_3$.

The LF analysis also reveals a very strong 1x1 modulation of the imaginary lattice Bragg peaks $\Im A_n^G$, while $\Re A_n^G$ remains smoother. To track this modulation, we define a two-dimensional nematic order parameter $(N_x, N_y) = \Im(\frac{-\sqrt{3}}{2}(A_2^G - A_3^G), A_1^G - A_2^G/2 - A_3^G/2)$, that vanishes above $T_{CDW}$ and selects a preferred axis as the CDW sets in. Figure 4e shows its modulated intensity and direction dependence across the single super-domain. Intriguingly, this nematicity is strongest in the domain walls, where the CDW is weakest. Moreover, $\bar{N}$ points along $\bar{G}_1$ for wide domain walls, while it points along $\bar{G}_3$ for narrow domain walls. This dependence, which is observed in all instances analyzed, shows nematicity is strongly coupled to the CDW modulation.

The origin of the observed domain wall phenomenology can be understood from the Ginzburg Landau theory of coupled CDW order parameters $\Delta_n$ and $\phi_n$, whose energy density can be written as $F = F_{CCDW} + F_\partial^\Delta + F_\partial^\phi + F_\partial^{\Delta\phi} + F_N$. The first term governs the homogenous CCDW phase and is given by

$$F_{CCDW} = \sum_n [a_\Delta \Delta_n^2 + c_\Delta \Delta_n^4 + a_\phi \phi_n^2 + b_\phi \phi_1 \phi_2 \phi_3 + c_\phi \phi_n^4 + b_{\Delta\phi} \phi_n \Delta_{n+1} \Delta_{n+2}],$$

where we have only included a single quartic term for simplicity. Finite primary $\Delta_n$ develop when $a_\phi < 0$, which immediately leads to finite secondary $\phi_n$ through a finite $b_{\Delta\phi}$ (even when $a_\phi > 0$). Because $\phi_n$ is mirror even, a cubic term $b_\phi$ is allowed (unlike for $\Delta_n$); this coefficient selects the sign of $\phi_1\phi_2\phi_3$. Indeed, we have experimentally observed only ground states where $\phi_1\phi_2\phi_3 > 0$ that fixes $b_\phi < 0$, whereas both signs of $\Delta_1\Delta_2\Delta_3$ are observed, in agreement with the absence of a cubic term for $\Delta_n$ in $F_{CCDW}$ (which imposes a degeneracy for both possibilities). The derivative terms which drive the incommensuration take the form

$$F_\partial^\Delta = \sum_{n,i} d_1^\Delta (\partial_i \Delta_n)^2 + d_2^\Delta \bar{Q}_n \cdot \left[2(\partial_x \Delta_n)(\partial_y \Delta_n), (\partial_x \Delta_n)^2 - (\partial_y \Delta_n)^2\right] + d_3^\Delta \left((\partial_x^2 + \partial_y^2)\Delta_n\right)^2$$

for $\Delta_n$, $F_\partial^\phi$ takes the same form as $F_\partial^\Delta$ replacing $\phi_n$ by $\Delta_n$, and a mixed term $F_\partial^{\Delta\phi} = \sum_n d_1^{\Delta\phi} \bar{Q}_n \times (\Delta_n \bar{\partial} \phi_n - \phi_n \bar{\partial} \Delta_n)$. Interestingly, we observe almost no dephasing between $\Delta_n$ and $\phi_n$, which suggests that the incommensuration is not driven by the linear derivative term $d_1^{\Delta\phi}$, contrary to what happens in McMillan theory of complex order parameters[3]. Rather, the fact that $\Delta_n$ and $\phi_n$ are locked together points to an instability driven by a negative quadratic derivative $d_1^\Delta < 0$ (with $d_3^\Delta > 0$ so that $F$ remains bounded), which is characteristic of Ising domain walls in incommensurate systems with real order parameters. Finally, the coupling to the nematic order parameter occurs via the linear coupling $F_N = a_N \left( \frac{-\sqrt{3}}{2} (\Delta_2^2 - \Delta_3^2), \Delta_1^2 - \Delta_2^2/2 - \Delta_3^2/2 \right) \cdot (N_x, N_y)$.

By minimizing the proposed free energy numerically according to the methods in Ref. [35], we indeed found a solution, shown in Fig. 4f, which qualitatively reproduces all experimental observations, shown in the same format in Fig. 4g for comparison. First, the harmonic content where one CDW order parameter component modulates with $2\bar{q}_{NC}$ while the other two modulate with $\bar{q}_{NC}$ and $3\bar{q}_{NC}$ is exactly reproduced, and is generated by the cubic coupling $b_{\Delta\phi}$ as well as $b_\phi$. This reveals the important role of the secondary order parameter in the NC state, since the primary order parameter alone does not have a cubic term and hence this type of modulated solution is impossible for $\Delta_n$ alone. Second, the nematic order parameter modulation $\bar{N}(\bar{q}_{NC}\bar{x})$ (red arrows in Figs. 4f,g) is also reproduced. Inside domains, the local $\Delta_n$ configuration is approximately threefold symmetric, and the coupling to nematicity vanishes. However, at domain walls, two components of $\Delta_n$ cross zero, while one remains finite, and the coupling in $F_N$ is maximal when $\bar{N}$ points along the corresponding $G_n$. This produces the alternation of the nematic direction in consecutive domain walls. Third, the existence of wide and narrow domain walls (Fig.3d) is explained from a small anisotropy coupling $d_2^\Delta$ and the fact that $\bar{q}_{NC}$ is not exactly parallel to $\bar{G}_2$ by a small angle. Because of this misalignment, $d_2^\Delta$ gives different energies to domain walls with nematicity pointing along $\bar{G}_1$ vs $\bar{G}_3$ naturally making one type of domain wider and the other narrower.

**Discussion**

By combining a rigorous symmetry analysis with phase locking techniques, our work has established the long-sought structure of the near-commensurate state in monolayer TiSe$_2$, revealing a number of surprising features. The NC state is unidirectional, and formed by the modulation of two intertwined real CDW order parameters with Ising domain walls, in accordance with its parent 2x2 CDW state. The free energy analysis also shows the coupling to the secondary order parameter to be crucial to obtain the observed NC states, as the established primary order parameter of TiSe$_2$ would not support such states alone. Finally, the NC state displays a strong nematic coupling at domain walls reflected in the lattice Bragg peaks.

It is important to note that while we have described the local domains as approximately threefold symmetric, the fact that NC super-domains are unidirectional implies that a small amount of anisotropy is naturally expected also within local domains. Because of this, our data on the NC state does not determine whether the parent commensurate state is threefold symmetric or not[36,37]. Nematicity only occurs within domain walls, and is probably not expected in the undoped, globally commensurate state. Similarly, while in our monolayer TiSe$_2$ samples inversion symmetry is broken by the $M_1^-$ CDW, this has no implications on whether bulk TiSe$_2$ with an $L_1^-$ CDW preserves this symmetry, which can only occur due to the admixture of extra order parameters[38–41]. The precise knowledge of the NC state also sets strong constraints for the interplay of CDW and superconductivity in TiSe$_2$. For example, the observed Little-Parks effect in the superconducting state was believed to be induced by a two-dimensional network of domain walls through which flux can be threaded[12], but it is unclear how this may happen for one dimensional networks, suggesting a different mechanism may be at play. Finally, by unveiling the nature of the NC state, our work more generally provides a key piece to understand the global phase diagram of period doubling CDWs, including their I-C transitions, emphasizing the importance of Ising domain walls and the difference with the usual NC states observed in complex CDWs with $N > 2$.

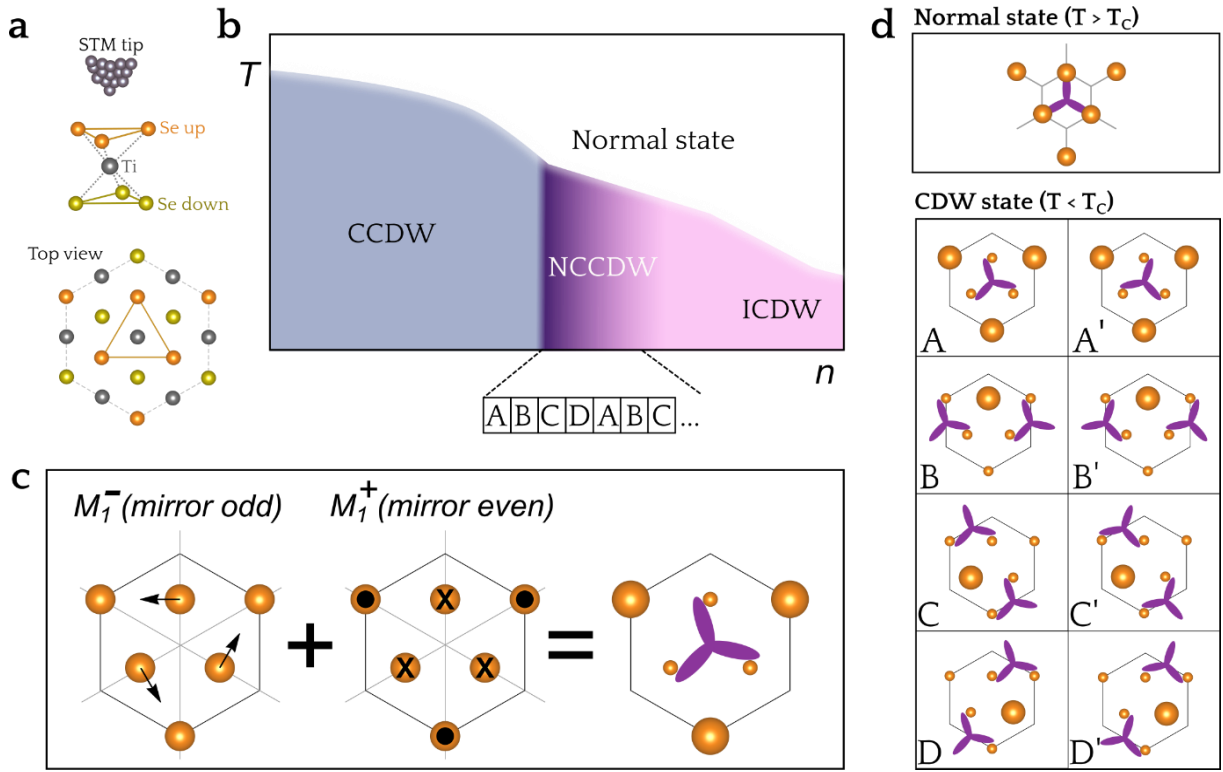

**Fig. 1 | CCDW and NCCDW in monolayer 1T-TiSe₂. a**, Side- and top-view of the crystal structure of monolayer TiSe$_2$. **b**, Schematic phase diagram of the doping (*n*)-induced nearly commensurate (NC) CDW transition in monolayer TiSe$_2$. **c**, Sketch of the displacements of the upper Se atoms corresponding to primary ($M_1^-$, mirror odd) and secondary ($M_1^+$, mirror even) order parameters of the transition, with the three mirror planes shown as thin gray lines. The sketch on the right is the resulting image commonly obtained in STM, with one brighter and three dimmer upper Se atoms, and a slightly rotated propeller-like feature. **d**, The eight possible ground states of the 2x2 CCDW. The four unprimed (ABCD) states represent four different choices of origin in the 2x2 unit cell, while the primed states are the same but with opposite rotation of the propeller. The structure of the NC state is shown below the phase diagram, as a one dimensional train of ABCD domains, with each domain spanning around 20 nm (or ~30 CDW unit cells).

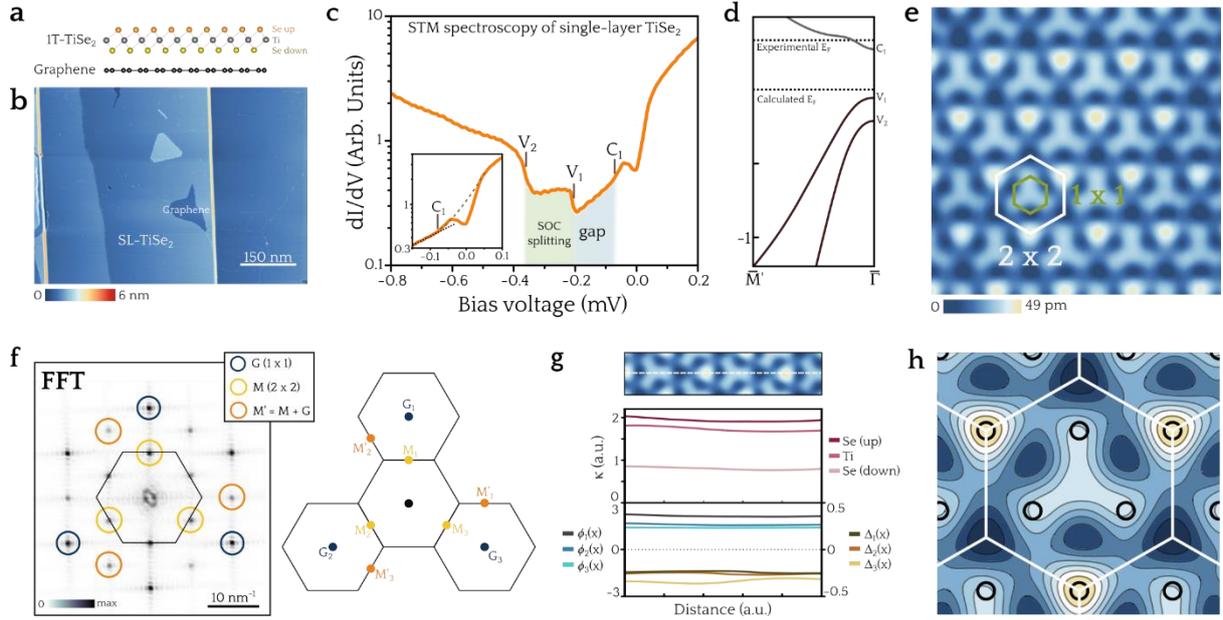

**Fig. 2 | Characterization of monolayer TiSe$_2$ on graphene and extraction of order parameters. a**, Side-view sketch of monolayer 1T-TiSe$_2$ on graphene. **b**, Large-scale STM image of nearly one layer of TiSe$_2$ on BLG/SiC(0001) ($V_s$ = 2.4 V, $I_t$ = 15 pA). **c**, Typical STM d$I$/d$V$ spectrum acquired on monolayer TiSe$_2$/BLG showing electronic features V$_{1-2}$ and C$_1$ ($I_t$ = 2 nA). The colored shadows below the curve indicate the SOC splitting (green) and the gap (blue). The inset shows a zoom around the C$_1$ feature. **d**, Calculated band structure of the isolated monolayer 1T-TiSe$_2$ in the CDW 2x2 phase (DFT-HSE hybrid functional) adapted from ref. [13]. The position of the experimental $E_F$ is also indicated. **e**, Atomically resolved STM image of monolayer 1T-TiSe$_2$ in the CDW state showing the 2x2 reconstruction ($V_s$ = 0.1 V, $I_t$ = 1.5 nA). The atomic and CDW unit cells are indicated in green and white, respectively. **f**, FFT of the STM image in **e**, showing CDW Bragg peaks at $M$ and $M'$ as defined in the right panel. **g**, Modulus $\kappa = \sqrt{[1/3 \sum_n (\phi_n^\alpha(x))^2]}$ for $\alpha$ = Se-Up, Ti, Se-down (upper plot), and extracted order parameters $\Delta_n(x)$ and $\phi_n(x) \equiv \phi_n^{SeUp}(x)$ from the CDW peak (lower plot) plotted along the white line in the zoomed-in region shown above. **j**, Real-space contour plot of $\rho(\bar{x})$ reconstructed by taking averaged values of the order parameters showing the slightly counter-clockwise rotated propeller when $\Delta_1\Delta_2\Delta_3 < 0$.

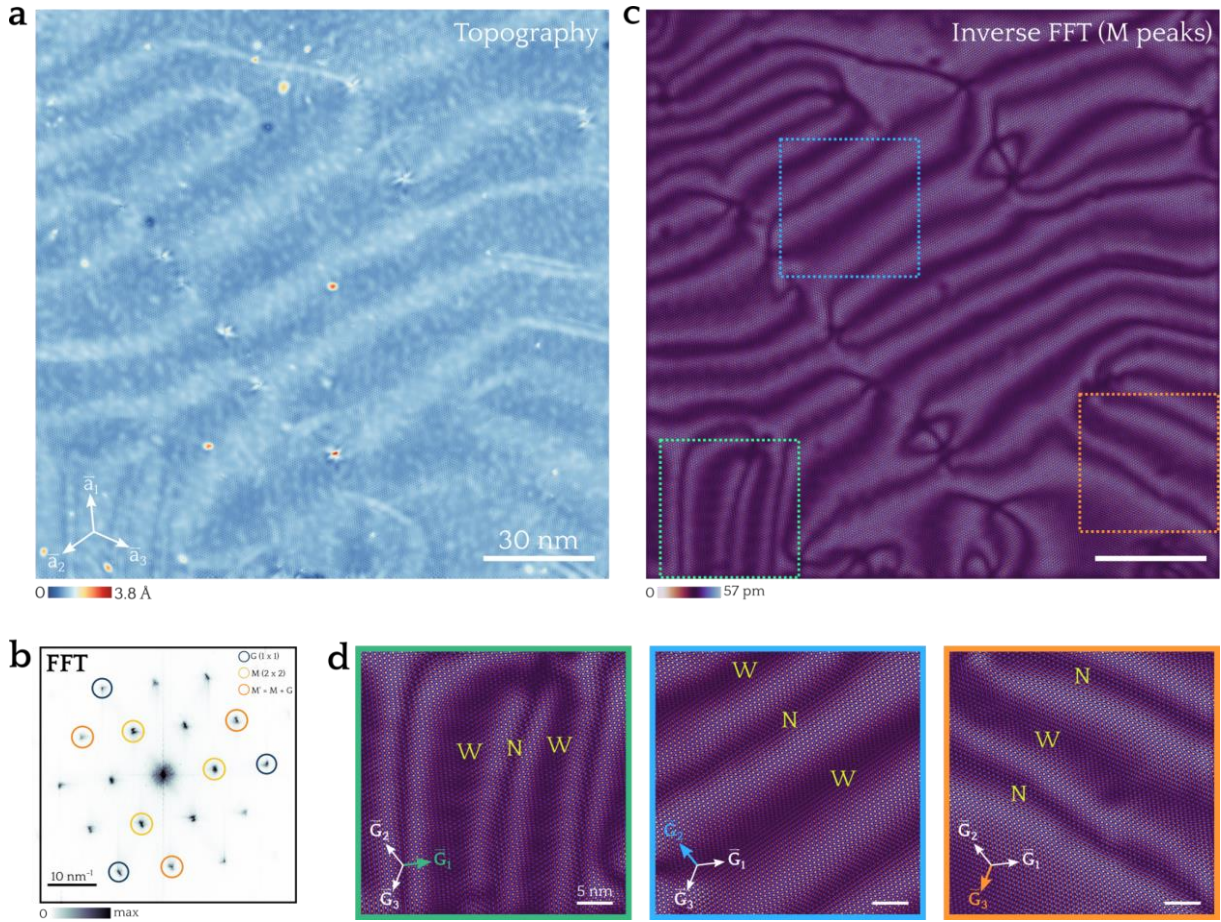

**Fig. 3 | Near-commensurate CDW state in monolayer TiSe$_2$. a**, Atomically resolved STM topography of a 153 nm × 153 nm region of single-layer TiSe$_2$ ($V_s$ = -0.05 V, $I_t$ = 60 pA). The white arrows are the lattice vectors. **b**, Corresponding Fourier transform of **a**. **c**, Inverse Fourier transform of the amplitude of the CDW M points from **a** (Supplementary Fig. S10 for *M'*). An intricate network of one dimensional trains of domains (high amplitude) separated by domain walls (low amplitude) with average spacing of 20 nm is observed. **d,** Zoom-in of three smaller regions shown in colored squares in **c**, which show domain walls are approximately perpendicular to the reciprocal lattice vectors G. The zoom also shows a finer structure: domain walls alternate between wide (W) and narrow (N) types.

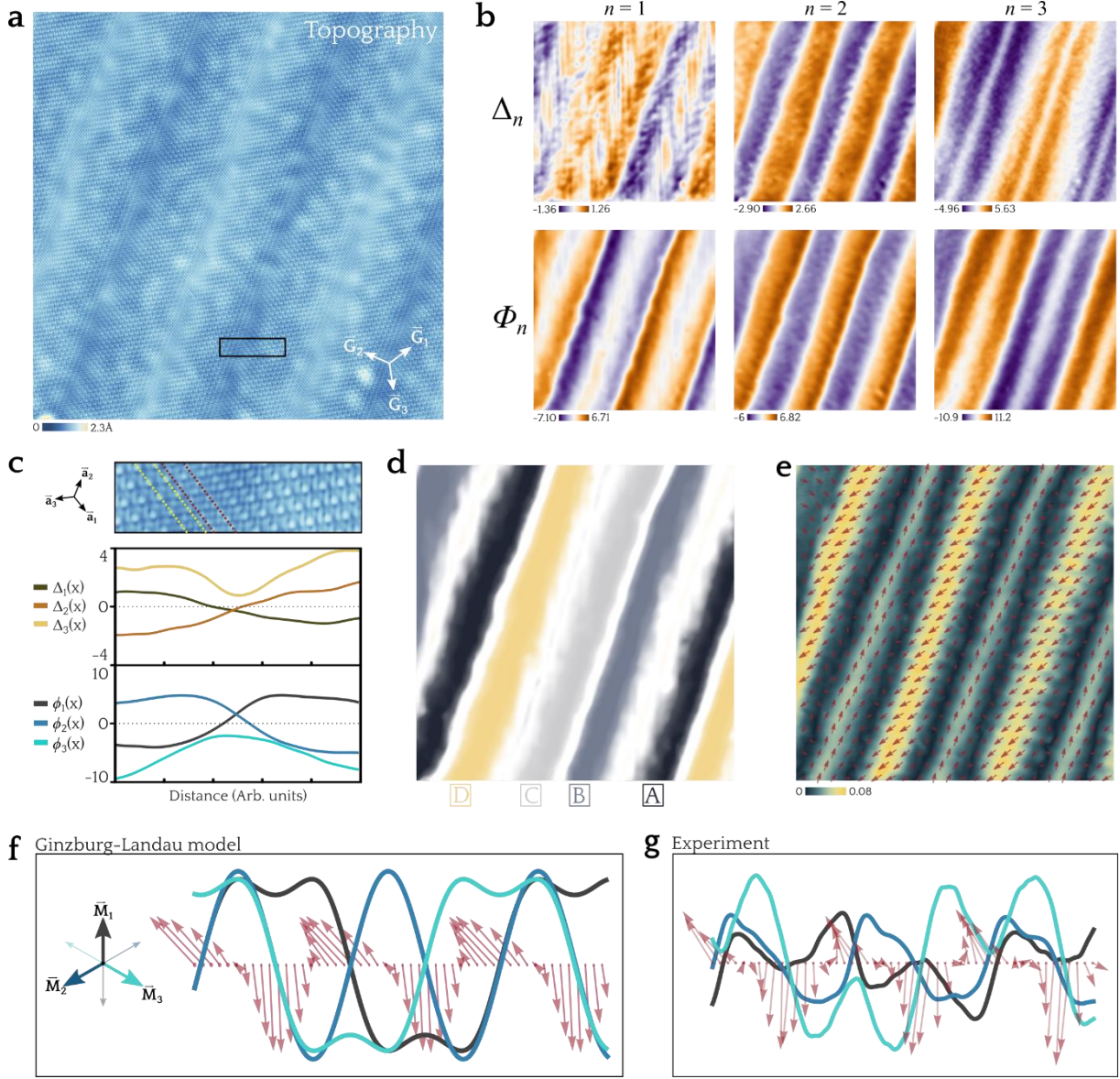

**Fig. 4 | Ising domains of intertwined CDW order parameters. a,** High-resolution topograph of a region showing domain walls after LF correction ($V_s$ = -0.05 V, $I_t$ = 40 pA). **b,** Extracted order parameters $\phi_n(\bar{x})$ and $\Delta_n(\bar{x})$. $\Delta_2$ oscillates twice as fast as $\Delta_{1,3}$ and that $\phi_n \sim -\Delta_n$. **c,** Close-up of a domain wall in the STM image in **a**, and corresponding one-dimensional cuts of the order parameters. The domain wall connects states C (left) to B (right), according to the convention in Fig. 1c. This corresponds to a real space shift by $\bar{a}_3$ (illustrated by dashed lines). **d,** Color map of domains extracted from **b**, which shows the train ABCD. **e**, Nematicity map $\bar{N}(\bar{x})$ obtained from the lattice Bragg peaks. The color maps shows the modulus of $N$, while the arrows show its direction. $\bar{N}(\bar{x})$ lives on domain walls, and alternates its direction between $\bar{G}_1$ (wide domain walls) and $\bar{G}_3$ (narrow domain walls). A one-dimensional cut of $\phi_n(\bar{q}_{NC}\bar{x})$ (full lines) and $\bar{N}(\bar{q}_{NC}\bar{x})$ (red arrows) from f, the Ginzburg-Landau prediction and g, STM experiment.

**References**

1. Chen, C.-W., Choe, J. & Morosan, E. Charge density waves in strongly correlated electron systems. *Reports Prog. Phys.* **79**, 084505 (2016).
2. Bruce, A. D., Cowley, R. A. & Murray, A. F. The theory of structurally incommensurate systems. II. Commensurate-incommensurate phase transitions. *J. Phys. C Solid State Phys.* **11**, 3591–3608 (1978).
3. McMillan, W. L. Theory of discommensurations and the commensurate-incommensurate charge-density-wave phase transition. *Phys. Rev. B* **14**, 1496 (1976).
4. Kogar, A. *et al.* Observation of a Charge Density Wave Incommensuration Near the Superconducting Dome in $Cu_xTiSe_2$. *Phys. Rev. Lett.* **118**, 027002 (2017).
5. Joe, Y. I. *et al.* Emergence of charge density wave domain walls above the superconducting dome in $1T-TiSe_2$. *Nat. Phys.* **10**, 421–425 (2014).
6. Morosan, E. *et al.* Superconductivity in $Cu_xTiSe_2$. *Nat. Phys.* **2**, 544–550 (2006).
7. Iavarone, M. *et al.* Evolution of the charge density wave state in $Cu_xTiSe_2$. *Phys. Rev. B* **85**, 155103 (2012).
8. Novello, A. M. *et al.* Stripe and Short Range Order in the Charge Density Wave of $1T-Cu_xTiSe_2$. *Phys. Rev. Lett.* **118**, 017002 (2017).
9. Yan, S. *et al.* Influence of Domain Walls in the Incommensurate Charge Density Wave State of Cu Intercalated $1T-TiSe_2$. *Phys. Rev. Lett.* **118**, 106405 (2017).
10. Spera, M., Scarfato, A., Giannini, E. & Renner, C. Energy-dependent spatial texturing of charge order in $1T-Cu_xTiSe_2$. *Phys. Rev. B* **99**, 155133 (2019).
11. Lee, K. *et al.* Metal-to-insulator transition in Pt-doped $TiSe_2$ driven by emergent network of narrow transport channels. *npj Quantum Mater.* **6**, 1–8 (2021).
12. Li, L. J. *et al.* Controlling many-body states by the electric-field effect in a two-dimensional material. *Nature* **529**, 185–189 (2016).
13. Chen, P. *et al.* Charge density wave transition in single-layer titanium diselenide. *Nat. Commun.* **6**, 8943 (2015).
14. Peng, J. P. *et al.* Molecular beam epitaxy growth and scanning tunneling microscopy study of $TiSe_2$ ultrathin films. *Phys. Rev. B* **91**, 121113 (2015).
15. Chen, P. *et al.* Dimensional Effects on the Charge Density Waves in Ultrathin Films of $TiSe_2$. *Nano Lett.* **16**, 6331–6336 (2016).
16. Sugawara, K. *et al.* Unconventional charge-density-wave transition in monolayer $1T-TiSe_2$. *ACS Nano* **10**, 1341–1345 (2016).
17. Britnell, L. *et al.* Strong light-matter interactions in heterostructures of atomically thin films. *Science* **340**, 1311–1314 (2013).
18. Fang, X. Y., Hong, H., Chen, P. & Chiang, T. C. X-ray study of the charge-density-wave transition in single-layer $TiSe_2$. *Phys. Rev. B* **95**, 201409 (2017).
19. Watson, M. D. *et al.* Strong-coupling charge density wave in monolayer $TiSe_2$. *2D Mater.* **8**, 015004 (2020).
20. Singh, B., Hsu, C. H., Tsai, W. F., Pereira, V. M. & Lin, H. Stable charge density wave phase in a $1T-TiSe_2$ monolayer. *Phys. Rev. B* **95**, 245136 (2017).
21. Chen, C., Singh, B., Lin, H. & Pereira, V. M. Reproduction of the Charge Density Wave Phase Diagram in $1T-TiSe_2$ Exposes its Excitonic Character. *Phys. Rev. Lett.* **121**, 226602 (2018).
22. Wei, M. J. *et al.* Manipulating charge density wave order in monolayer $1T-TiSe_2$ by strain and charge doping: A first-principles investigation. *Phys. Rev. B* **96**, 165404 (2017).
23. Guster, B., Canadell, E., Pruneda, M. & Ordejón, P. First principles analysis of the CDW instability of single-layer $1T-TiSe_2$ and its evolution with charge carrier density. *2D Mater.* **5**,

025024 (2018).

24. Pasquier, D. & Yazyev, O. V. Excitonic effects in two-dimensional TiSe$_2$ from hybrid density functional theory. *Phys. Rev. B* **98**, 235106 (2018).
25. Zhou, J. S. *et al.* Anharmonicity and Doping Melt the Charge Density Wave in Single-Layer TiSe$_2$. *Nano Lett.* **20**, 4809–4815 (2020).
26. Ishioka, J. *et al.* Chiral charge-density waves. *Phys. Rev. Lett.* **105**, 176401 (2010).
27. Novello, A. M. *et al.* Scanning tunneling microscopy of the charge density wave in TiSe$_2$ in the presence of single atom defects. *Phys. Rev. B* **92**, 081101 (2015).
28. Hildebrand, B. *et al.* Local Real-Space View of the Achiral 1T-TiSe$_2$ 2×2×2 Charge Density Wave. *Phys. Rev. Lett.* **120**, 136404 (2018).
29. Kim, H., Jin, K. H. & Yeom, H. W. Electronically Seamless Domain Wall of Chiral Charge Density Wave in 1T-TiSe$_2$. *Nano Lett.* **24**, 14323–14328 (2024).
30. Di Salvo, F. J., Moncton, D. E. & Waszczak, J. V. Electronic properties and superlattice formation in the semimetal TiSe$_2$. *Phys. Rev. B* **14**, 4321 (1976).
31. Ueda, H. *et al.* Correlation between electronic and structural orders in 1T-TiSe$_2$. *Phys. Rev. Res.* **3**, L022003 (2021).
32. Wickramaratne, D., Subedi, S., Torchinsky, D. H., Karapetrov, G. & Mazin, I. I. Photoinduced chiral charge density wave in TiSe$_2$. *Phys. Rev. B* **105**, 054102 (2022).
33. Chen, C., Su, L., Castro Neto, A. H. & Pereira, V. M. Discommensuration-driven superconductivity in the charge density wave phases of transition-metal dichalcogenides. *Phys. Rev. B* **99**, 121108 (2019).
34. Novello, A. M. *et al.* Scanning tunneling microscopy of the charge density wave in 1T-TiSe$_2$ in the presence of single atom defects. *Phys. Rev. B* **92**, 081101 (2015).
35. Nakanishi, K. & Shiba, H. Domain-Like Incommensurate Charge-Density-Wave States and the First-Order Incommensurate-Commensurate Transitions in Layered Tantalum Dichalcogenides. I. 1T-Polytype. *J. Phys. Soc. Japan* **43**, 1839–1847 (1977).
36. Subedi, A. Trigonal-to-monoclinic structural transition in TiSe$_2$ due to a combined condensation of q=(1/2,0,0) and (1/2,0,1/2) phonon instabilities. *Phys. Rev. Mater.* **6**, 014602 (2022).
37. Muñoz-Segovia, D., Venderbos, J. W. F., Grushin, A. G. & de Juan, F. Nematic and stripe orders within the charge density wave state of doped TiSe$_2$. *Arxiv* 2308.15541 (2023).
38. Xiao, Q. *et al.* Observation of giant circular dichroism induced by electronic chirality. *Phys. Rev. Lett.* **133**, 126402 (2023).
39. Kim, K. *et al.* Origin of the chiral charge density wave in transition-metal dichalcogenide. *Nat. Phys.* 10.1038/s41567-024-02668-w (2024).
40. Peng, Y. *et al.* Observation of orbital order in the van der Waals material 1T-TiSe$_2$. *Phys. Rev. Res.* **4**, 033053 (2022).
41. Xu, S. Y. *et al.* Spontaneous gyrotropic electronic order in a transition-metal dichalcogenide. *Nature* **578**, 545–549 (2020).
42. Horcas, I. *et al.* WSxM: A software for scanning probe microscopy and a tool for nanotechnology. *Rev. Sci. Instrum.* **78**, 013705 (2007).
43. Lawler, M. J. *et al.* Intra-unit-cell electronic nematicity of the high-Tc copper-oxide pseudogap states. *Nature* **466**, 347–351 (2010).

## Methods

**Sample preparation and initial characterization of single-layer TiSe₂**

Single-layer TiSe$_2$ samples were epitaxially grown on bilayer graphene (BLG) on SiC(0001) substrates. First, uniform bilayer graphene was prepared by direct annealing 6H-SiC (0001) at a temperature around 1400 °C for 35 min. For the growth of monolayer TiSe$_2$, we co-evaporated high-purity Ti (99.95%) and Se (99.999%) in our home-made molecular beam epitaxy (MBE) system under base pressure of ~5 ×10$^{-10}$ mbar. The flux ratio between Ti and Se is 1:30. During the growth, the temperature of BLG/SiC(0001) substrates were maintained at 450°C and the growth rate was ~20 minutes/monolayer. After the growth of monolayer TiSe$_2$, the samples were kept annealed in the Se environment for 2 minutes to minimize the presence of atomic vacancies, and then immediately cooled down to room temperature. In-situ RHEED was used for monitoring the growth process. Atomic Force Microscopy at ambient conditions was routinely used to optimize the morphology, domain sizes, coverage and cleanliness of the TiSe$_2$ islands (see Supplementary Fig. 1). The samples used for AFM characterization were not further used for STM measurements. For further transfer of the samples from our MBE to the UHV-STM chamber, a Se capping layer with a thickness of ~10 nm was deposited on the surface to protect the film from contamination during transport. The Se capping layer was subsequently removed in the UHV-STM by annealing the sample at 300°C for 40 minutes prior to the STM measurements.

**STM/STS measurements**

STM/STS measurements were performed on a commercial STM (USM1300, Unisoku) operated at 4.2 K using Pt/Ir tips, which were calibrated on Au(111) prior to the STM/STS measurements. Standard lock-in technique was employed for STS data acquisition, using AC modulation voltages $V_{a.c.}$ ~1 mV at $f$ = 833 Hz. All STM/STS data were post-processed and analyzed employing the freeware WSxM[42].

**Phase-locking analysis**

To extract the primary and secondary order parameters defined in the main text, we need the spatially dependent complex amplitudes $A_n^M(\bar{x})$ and $A_n^{M'}(\bar{x})$ with well defined phases so that the real and imaginary parts can be separated. For this, we use geometric phase analysis (https://github.com/TAdeJong/pyGPA) and implement the LF algorithm[43] to produce corrected STM images in perfect registry with the lattice, which enforces a constant phase of the complex numbers $A_n^G(\bar{x}) \equiv A_n^G$. To fix the origin we note there are three points in the unit cell where $arg A_n^G$ are approximately equal, which correspond to Ti, Se-Up and Se-Down atoms. We choose by convention that $-2\pi/3 < arg\ A_n^G < -\pi/3$ which selects the Ti site as origin and the order Se-Down-Ti-Se-Up as we move along (0,1) (see Fig. 1a). The $arg\ A_n^G(\bar{x})$ after the implementation of LF to the STM image of Fig.2e is shown in Fig.S6, which shows a constant value slightly bigger than $-2/3\pi$ for all three components. The corresponding 1x1 modulation is shown in Fig. S6b (see also Fig. 1d). The extracted 2x2 order parameters for this image are shown in Fig. 2g. Together with the extracted amplitude of the 1x1 lattice peaks ($|A_n^G| = 2.72$) one can build the total charge density $\rho(\bar{x}) = \rho_{1x1}(\bar{x}) + \rho_{2x2}(\bar{x})$, illustrated in Fig. 2h. The same procedure was used in Fig. 4.

## Acknowledgements

M.M.U. acknowledges support by the ERC Starting grant LINKSPM (Grant 758558) and by the Spanish grant no. PID2020-116619GB-C21 funded by MCIN/AEI/10.13039/501100011033. F. J. acknowledges support from grant PID2021-128760NB0-I00 from the same institution. M.N.G. is supported by the Ramon y Cajal Fellowship RYC2021-031639-I. D.M.S. acknowledges support from from the Spanish MCIU FPU fellowship No. FPU19/03195.

## Author contributions

M.M.U. and F.J. conceived the project. W.W. and P.D. carried out the growth and subsequent STM/STS measurements. W.W. analyzed the STM/STS data under the supervision of M.M.U.. M.N.G. performed the data analysis and order parameter extraction with phase locking methods. M.N.G. and F.J. developed the Ginzburg-Landau analysis and theoretical interpretation of the data with support from D.M.S.. M.N.G., M.M.U. and F.J. wrote the manuscript. All authors contributed to the scientific discussion and manuscript revisions.

## Data availability

The data that support the findings of this study are available from the corresponding author upon reasonable request.

## Competing financial interests

The authors declare no competing financial interests.

Supplementary Materials for

# Ising domain wall networks from intertwined charge density waves in single-layer TiSe$_2$

Wen Wan[1], Maria N. Gastiasoro[1], Daniel Muñoz-Segovia[1], Paul Dreher[1], Miguel M. Ugeda*,[1,2,3] and Fernando de Juan*,[1,3]

[1]*Donostia International Physics Center, Paseo Manuel de Lardizábal 4, 20018 Donostia-San Sebastián, Spain*

[2]*Centro de Física de Materiales, Paseo Manuel de Lardizábal 5, 20018 San Sebastián, Spain.*

[3]*Ikerbasque, Basque Foundation for Science, Plaza Euskadi 5, 48009 Bilbao, Spain*

*\*Corresponding authors: mmugeda@dipc.org, fernando.dejuan@dipc.org*

**This PDF file includes:**

1. Complex vs real CDW order parameters

2. Sample preparation and initial characterization of single-layer TiSe$_2$

3. Large-scale electronic structure of single-layer TiSe$_2$

4. Primary and secondary order parameters of the CDW

5. STM imaging of the CDW pattern

6. Magnetic-field insensitivity of the CDW pattern

7. Temperature dependence of the CDW patterns

8. Fourier-filtered STM imaging of relevant periodicities

## 1. Complex vs real CDW order parameters

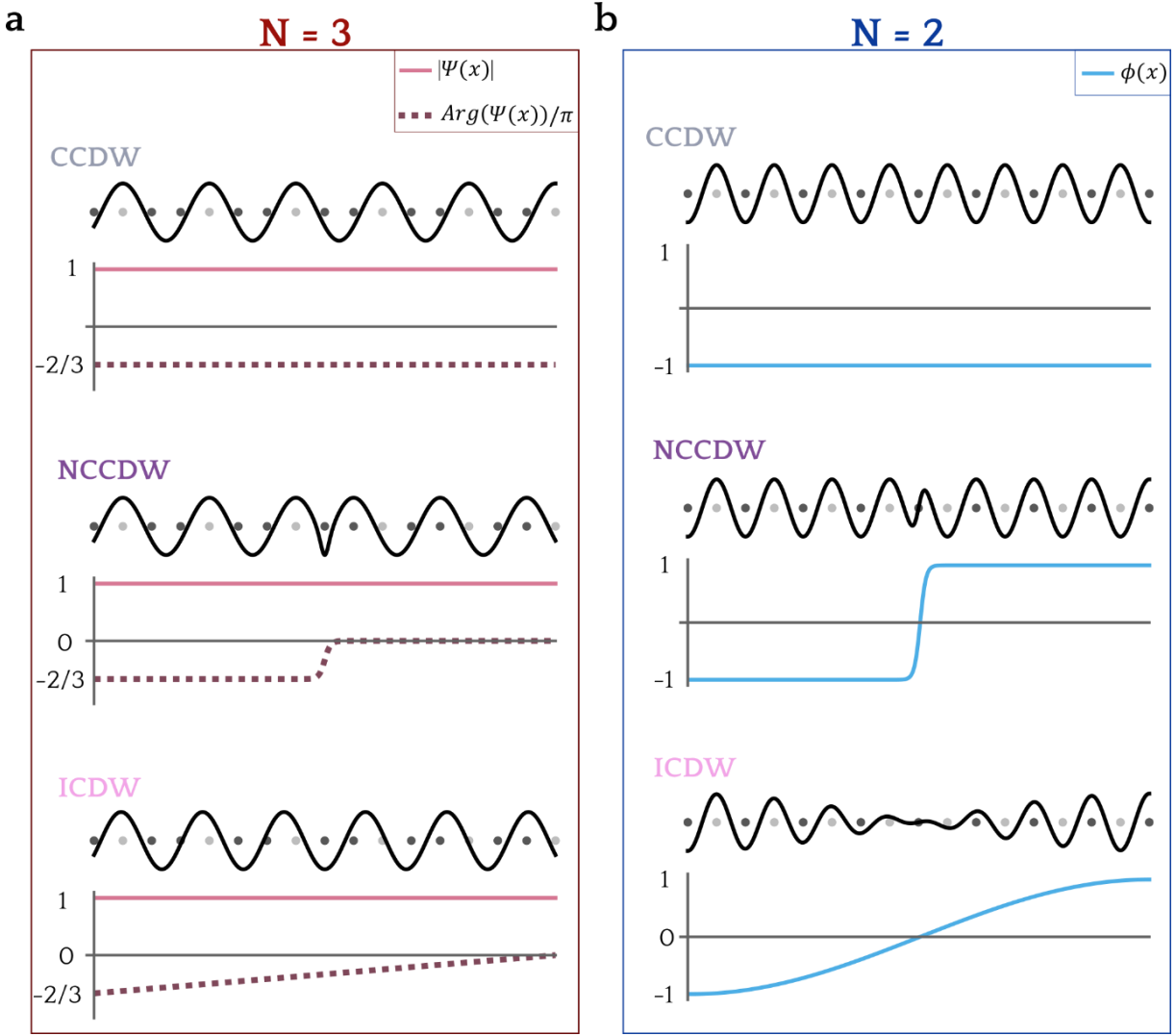

**Fig. S1 | Complex vs. real order parameters in C-NC-I CDW transitions with $Q = 2\pi/Na$. a**, Example of a 1D CDW with $Q = 2\pi/3a$ ($N = 3$), where the order parameter is complex and lattice translations are represented by a phase, $\Psi(x + a) = e^{iQa}\Psi(x) = e^{i2\pi/3}\Psi$. The top panel shows the commensurate state. An incommensurate state can be represented as a state where the phase grows linearly with constant amplitude (bottom panel). This state is unstable to a near-commensurate CDW (middle panel) where the phase grows via sharper phase slips known as discommensurations. **b**, A CDW with $Q = \pi/a$ ($N = 2$), where the real and imaginary parts of the Fourier transformed charge density represent independent real order parameters with different symmetry. The top panel shows the CCDW for the real part $\phi$. In this case lattice translations are represented by a sign change $\phi(x + a) = e^{iQa}\phi(x) = -\phi(x)$. An incommensurate state (bottom panel) is described by a smoothly modulating $\phi$, and this state is unstable to an NC state where $\phi$ is locally constant within domains and has sharper sign changes between them known as Ising domain walls.

## 2. Sample preparation and initial characterization of single-layer TiSe$_2$

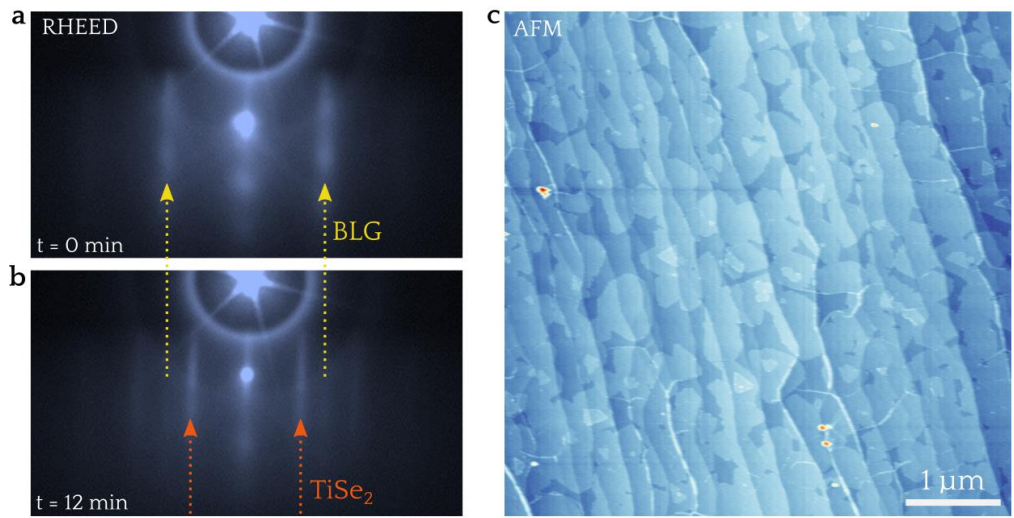

**Fig. S2 | Initial characterization of the TiSe$_2$ monolayers. a**, RHEED pattern after the growth of BLG/SiC(0001). Yellow lines indicate the diffraction features of BLG. The growth of the TiSe$_2$ monolayers gradually leads to new diffraction lines (red arrows) along with the attenuation of the BLG pattern, as shown in **b**. **c**, Amplitude-modulation AFM image showing the large-scale morphology of our single-layer TiSe$_2$ films on BLG/SiC(0001).

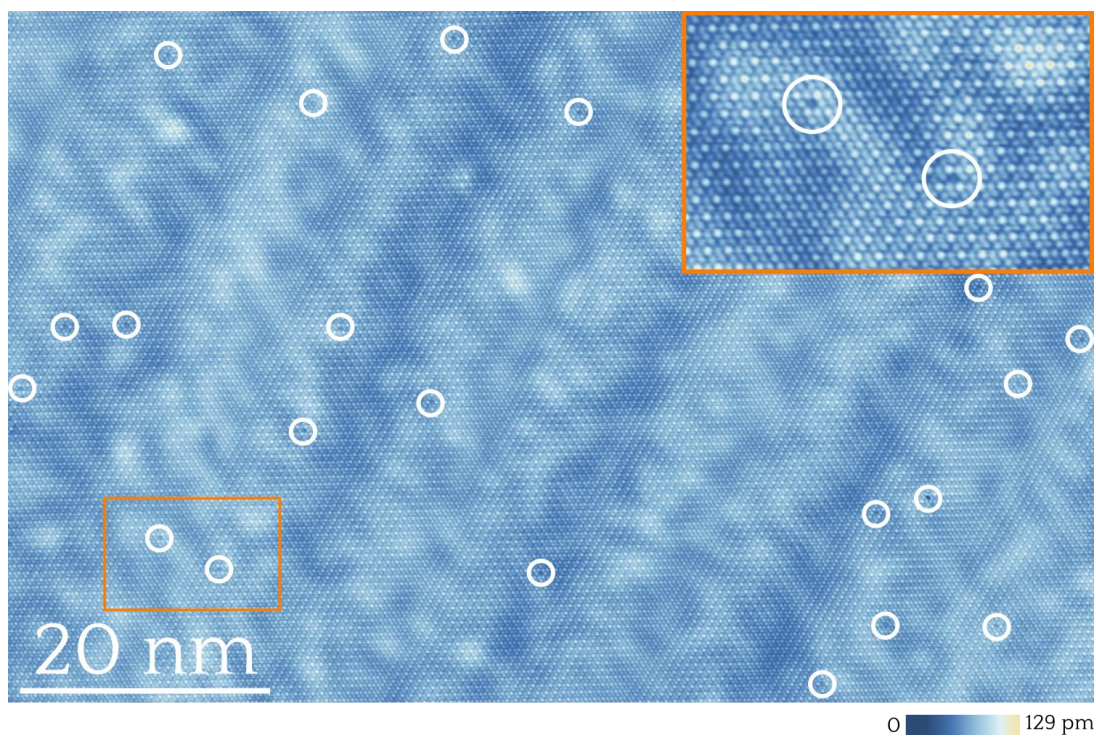

**Fig. S3 | Defect density in single-layer TiSe$_2$/BLG.** Typical atomically resolved STM image of our single-layer TiSe$_2$ films. White circles mark the position of point defects in the lattice. ($V_s$ = -0.05 V, $I_t$ = 60 pA). The inset shows a zoom in of the boxed region in the main image.

## 3. Large-scale electronic structure of single-layer TiSe$_2$

The electronic structure of single-layer TiSe$_2$ over a larger energy scale (± 2V) is shown in Fig. S4, compared to an *ab-initio* calculation of the single-layer band structure reproduced from ref. 1. The Y-axis (energy) between the two panels is rigidly shifted by 300 meV to account for the doping induced by the BLG substrate in the experiment. The d$I$/d$V$ spectrum shows three main features, which derive from features in the band structure which are unrelated to the CDW state. A rather sharp onset in the occupied states labeled as V$_3$ corresponds to tunneling into the Se p$_z$ orbitals, which in the monolayer occur at much lower energies than the bulk, as seen in the ab-initio calculation. Features C$_2$ and C$_3$ in the unoccupied states correspond to Van Hove singularities in the t$_{2g}$-like bands of Ti d orbitals, which are also present in bulk TiSe$_2$ (ref. 2).

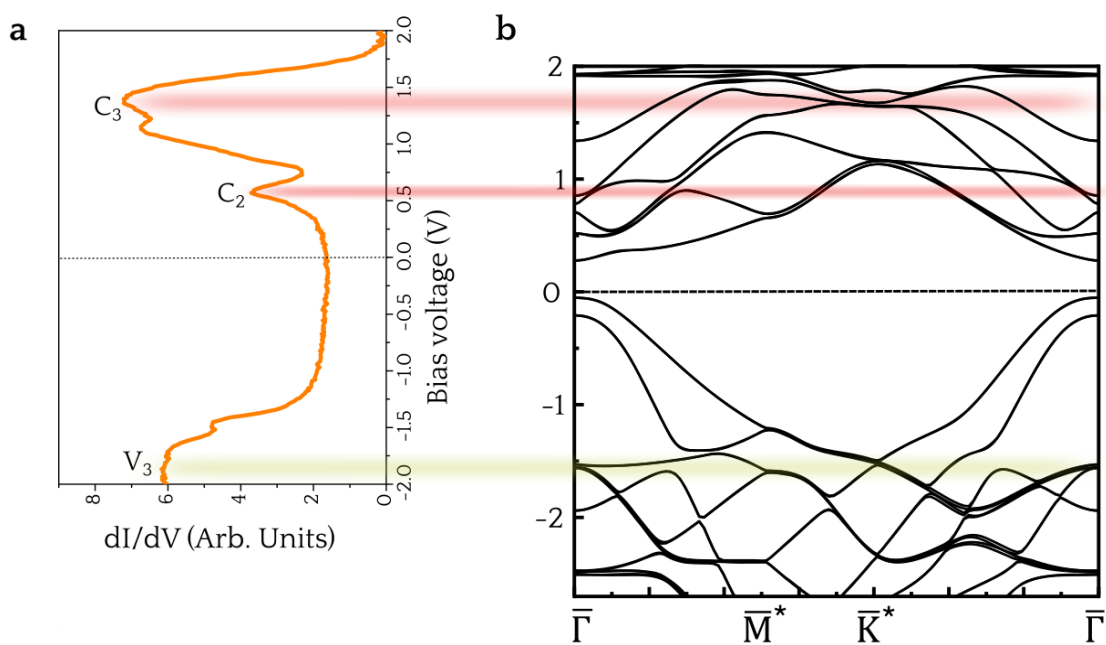

**Fig. S4 | Large-scale tunneling spectroscopy. a**, Representative large-scale d$I$/d$V$ curve acquired on single-layer TiSe$_2$/BLG ($T$ = 4.2 K). Further spectroscopic features labeled as V$_3$, C$_2$ and C$_3$ are shown, apart from those mentioned in the main manuscript (Fig. 2b) **b**, Calculated band structure for the monolayer TiSe$_2$ in the CDW state, adapted from ref. 1.

## 4. Primary and secondary order parameters of the CDW

The order parameters used in the main text are the primary order parameter $\Delta_n$ of $M_1^-$ and three secondary order parameters $\phi^{SeUp}$, $\phi^{Ti}$, $\phi^{SeDo}$ of symmetry $M_1^+$. These order parameters are extracted as follows. The charge density in the CDW state can be decomposed into the original 1x1 lattice modulation and the period doubling 2x2 modulation of the CDW, $\rho(\bar{x}) = \rho_{1x1}(\bar{x}) + \rho_{2x2}(\bar{x})$. We can express the former in terms of the complex numbers $A_n^G(\bar{x})$ at the $\bar{G}_n$ Bragg peaks of the FFT,

$$\rho_{1x1}(\bar{x}) = 2\Re \sum_n A_n^G(\bar{x}) e^{-i\bar{G}_n\bar{x}} = 2 \sum_n \Re A_n^G(\bar{x})\cos(\bar{G}_n\bar{x}) + \Im A_n^G(\bar{x})\sin(\bar{G}_n\bar{x}).$$

The Lawler-Fujita method (see details below) enforces the phase of $A_n^G(\bar{x})$ to be constant. The 2x2 part of the charge modulation can be equivalently expressed in terms of the complex numbers $A_n^M(\bar{x})$ and $A_n^{M'}(\bar{x})$ at the $\bar{M}_n$ and $\bar{M}_n' = \bar{M}_n + \bar{G}_{n-1}$ peaks of the FFT,

$$\rho_{2x2}(\bar{x}) = 2\Re \sum_n \left[ A_n^M(\bar{x}) e^{-i\bar{M}_n\bar{x}} + A_n^{M'}(\bar{x}) e^{-i\bar{M}_n'\bar{x}} \right]$$

$$= 2 \sum_n \Re A_n^M(\bar{x}) \cos(\bar{M}_n\bar{x}) + \Im A_n^M(\bar{x})\sin(\bar{M}_n\bar{x}) + \Re A_n^{M'}(\bar{x}) \cos(\bar{M}_n'\bar{x}) + \Im A_n^{M'}(\bar{x})\sin(\bar{M}_n'\bar{x})$$

$$\equiv 2 \sum_n \phi_n^{SeUp}(\bar{x}) f_n^{SeUp}(\bar{x}) + \phi_n^{Ti}(\bar{x}) f_n^{Ti}(\bar{x}) + \phi_n^{SeDo}(\bar{x}) f_n^{SeDo}(\bar{x}) + \Delta_n(\bar{x}) f_n^{\Delta}(\bar{x})$$

Here we have defined the secondary $M_1^+$ order parameters

$$\phi_n^{SeUp}(\bar{x}) = \Re A_n^{M'}(\bar{x}) - \frac{1}{2} \Re A_n^M(\bar{x}) + \frac{\sqrt{3}}{2} \Im A_n^M(\bar{x})$$

$$\phi_n^{Ti}(\bar{x}) = \Re A_n^{M'}(\bar{x}) + \Re A_n^M(\bar{x})$$

$$\phi_n^{SeDo}(\bar{x}) = \Re A_n^{M'}(\bar{x}) - \frac{1}{2} \Re A_n^M(\bar{x}) - \frac{\sqrt{3}}{2} \Im A_n^M(\bar{x})$$

with their corresponding site selective functions

$$f_n^{SeUp}(\bar{x}) = \frac{1}{3}\cos(\bar{M}_n'\bar{x}) - \frac{1}{3}\cos(\bar{M}_n\bar{x}) + \frac{1}{\sqrt{3}}\sin(\bar{M}_n\bar{x}) \quad (1)$$

$$f_n^{Ti}(\bar{x}) = \frac{1}{3}\cos(\bar{M}_n'\bar{x}) + \frac{2}{3}\cos(\bar{M}_n\bar{x}) \quad (2)$$

$$f_n^{SeDo}(\bar{x}) = \frac{1}{3}\cos(\bar{M}_n'\bar{x}) - \frac{1}{3}\cos(\bar{M}_n\bar{x}) - \frac{1}{\sqrt{3}}\sin(\bar{M}_n\bar{x}) \quad (3)$$

These site selective functions, shown in Figs. S5 b-d, are all mirror *even*, and can therefore only represent a 2x2 modulation of the secondary phonon of $M_1^+$ symmetry (1-in-4 pattern in Fig. 1c). The imaginary part of $A_n^{M'}(\bar{x})$, on the other hand, is identified with the primary CDW order parameter with symmetry $M_1^-$,

$$\Delta_n(\bar{x}) = \Im A_n^{M'}(\bar{x})$$

$$f_n^{\Delta}(\bar{x}) = \sin(\bar{M}_n'\bar{x}) \quad (4)$$

since it represents mirror *odd* in-plane modulations, illustrated in Fig. S5a.

Since all three $\phi^{SeUp}, \phi^{Ti}, \phi^{SeDo}$ have the same symmetry, it is sufficient to consider the one of largest magnitude for the Ginzburg-Landau analysis, since the other two will be locked to the first by a quadratic coupling. Because of this, in the main text we take $\phi_n = \phi_n^{SeUp}$. Lattice translations $a_n$ act on these order parameters with the operator $t_n = diag(e^{iG_1\bar{a}_n}, e^{iG_2\bar{a}_n}, e^{iG_3\bar{a}_n})$, which is explicitly $t_1 = diag(1,-1,-1)$, $t_2 = diag(-1,1,-1)$, $t_3 = diag(-1,-1,1)$. The different symmetry related ground states of the order parameters, shown in Fig. 1d, are explicitly defined as $\phi/\phi_0 = -\Delta/\Delta_0 = (1,1,1)$ for A, $\phi/\phi_0 = -\Delta/\Delta_0 = (1,-1,-1)$ for B, $\phi/\phi_0 = -\Delta/\Delta_0 = (-1,1,-1)$ for C and $\phi/\phi_0 = -\Delta/\Delta_0 = (-1,-1,1)$ for D. The primed states have the same $\phi$ but $\phi/\phi_0 = \Delta/\Delta_0$.

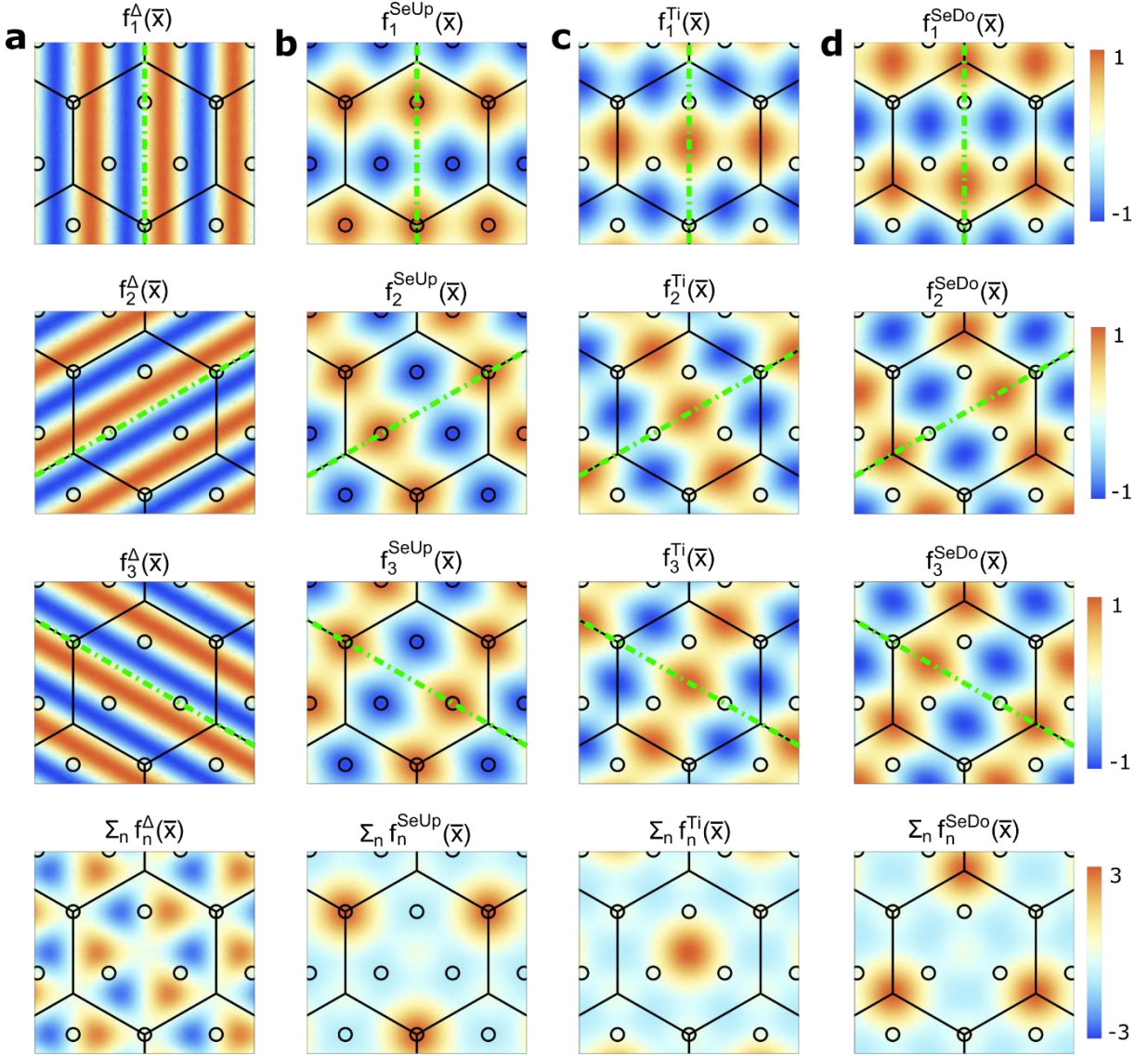

**Fig. S5 | Primary and secondary 2x2 order parameters. a**, Primary order parameter function components $f_n^\Delta(\bar{x})$ [eq. (4)]. **a**, Secondary order parameter site selective functions **b**, $f_n^{SeUp}(\bar{x})$ [eq. (1)] **c**, $f_n^{Ti}(\bar{x})$ [eq. (2)] and **d**, $f_n^{SeDo}(\bar{x})$ [eq. (3)]. The dotted lines represent the mirror $\sigma_v$ parallel to the corresponding $\bar{M}_n$ vector.

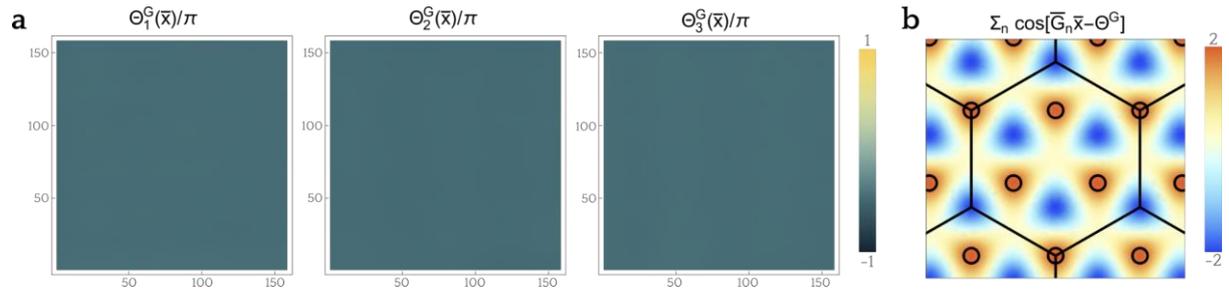

**Fig. S6 | Geometric phase analysis and LF algorithm. a,** Phase of the lattice Bragg peaks $\Theta_n^G(\bar{x}) = arg\, A_n^G(\bar{x})$ after implementation of the LF algorithm of an STM image from which Fig. 2e was extracted. It shows a constant value $\Theta_n^G(\bar{x}) = (-2/3 + 0.194)\pi$ throughout the field of view. **b,** Using the extracted phase $\Theta^G$ in a, the corresponding 1x1 lattice modulation $\rho_{1x1}(\bar{x})$ is reconstructed (see also normal state sketch in Fig. 1d).

## 5. STM imaging of the CDW pattern

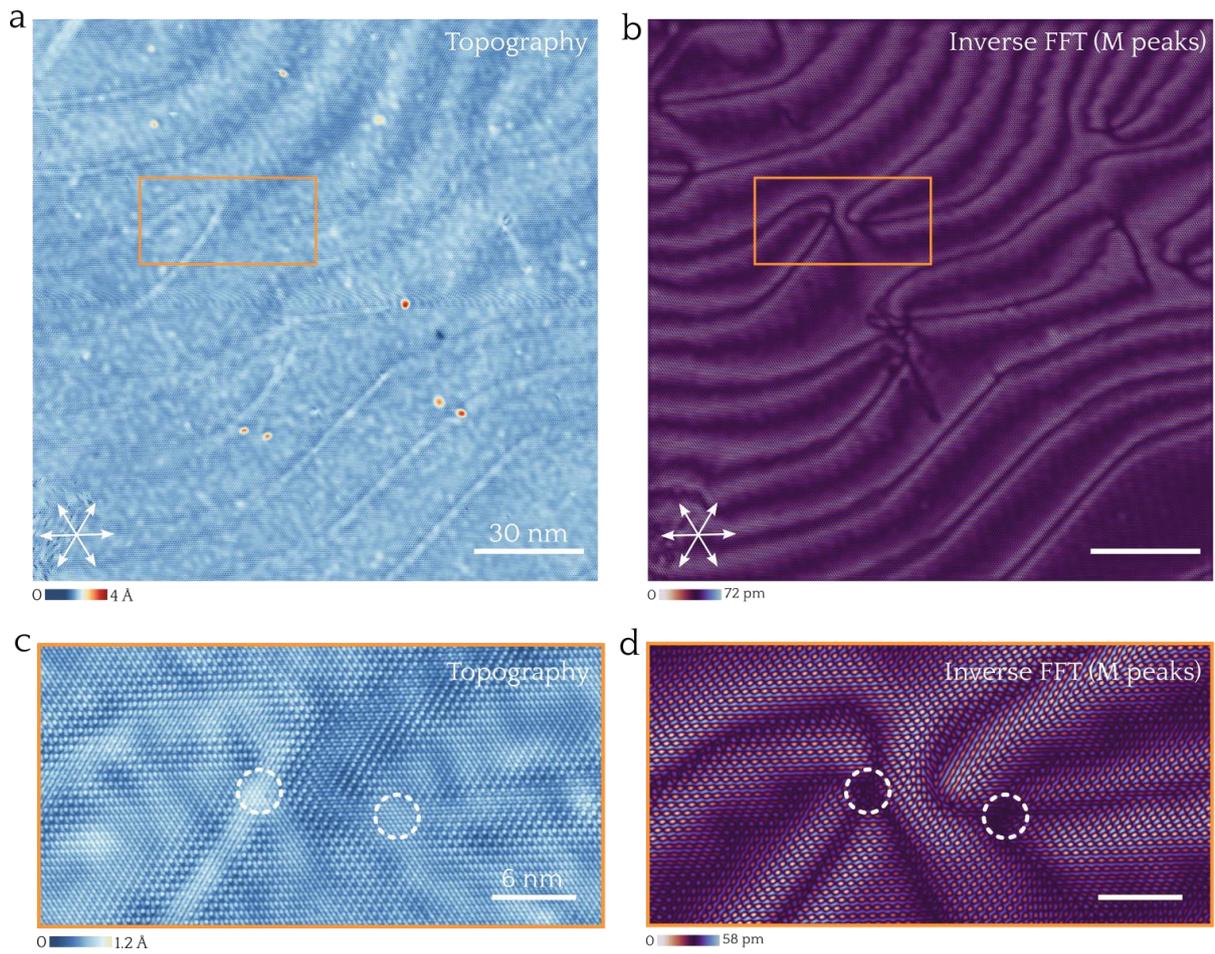

**Fig. S7 | Complex 2D pattern in the CDW state of single-layer TiSe$_2$.** **a** Atomically resolved STM topography of a 160 x 160 nm$^2$ region of single-layer TiSe$_2$ similar to tha shown in the main manuscript ($V_s$ = -0.05 V, $I_t$ = 50 pA). The white arrows indicate the crystal directions. **b** Inverse Fourier transform (FFT) of the CDW 2 x 2 peaks (M points) of the FFT (**c**) of the topography in **a**. **c** and **d** t shows zoomed-in images of the boxed region in the images in **a** and **b**, respectively.

## 6. Magnetic-field insensitivity of the CDW pattern

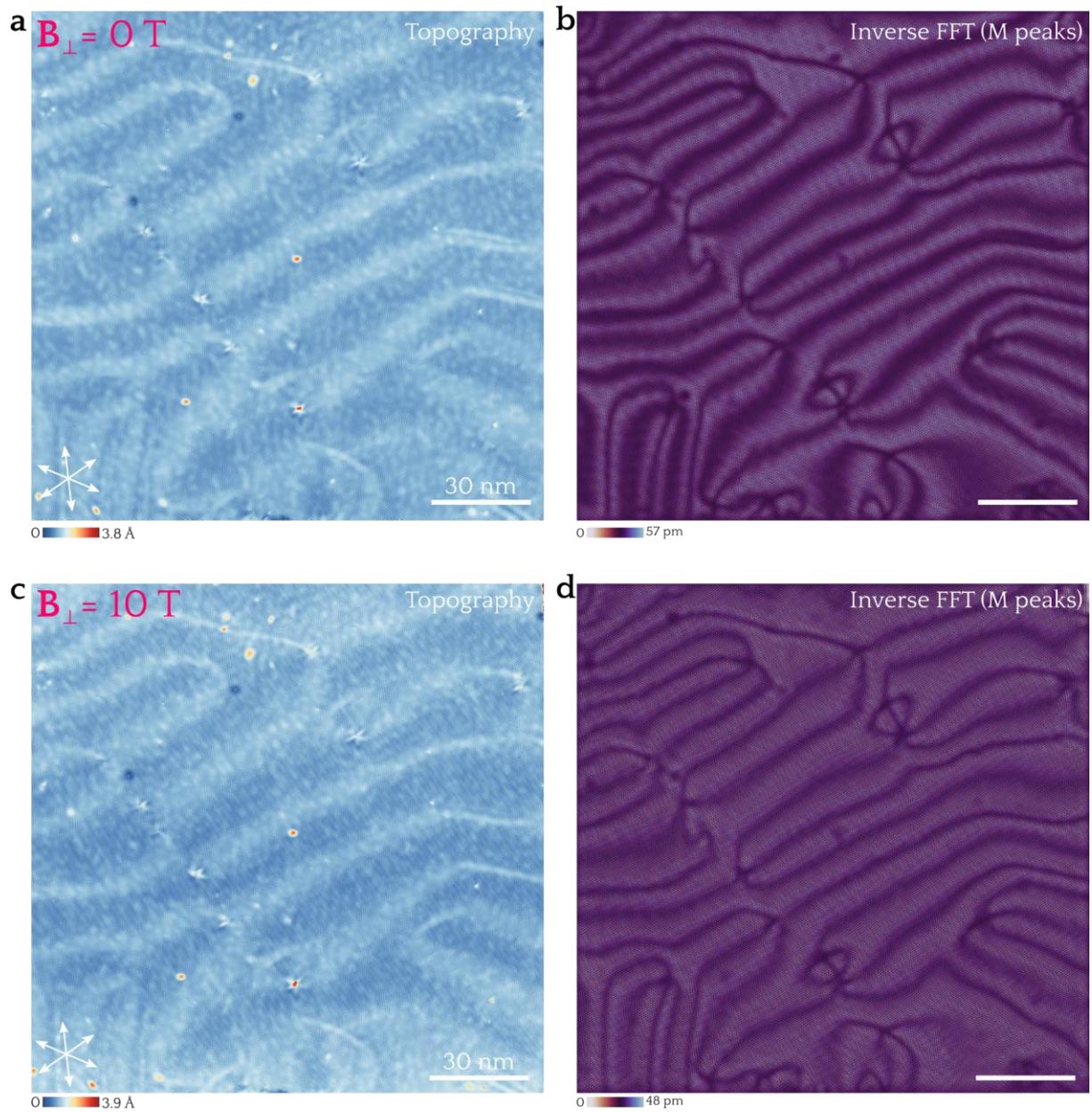

**Fig. S8 | Magnetic-field dependent STM imaging.** Atomically resolved STM topographs of single-layer TiSe$_2$ (**a**,**c**) and corresponding I-FFT of the CDW 2x2 peaks (**b**,**d**) acquired at different magnetic fields.

## 7. Temperature dependence of the CDW patterns

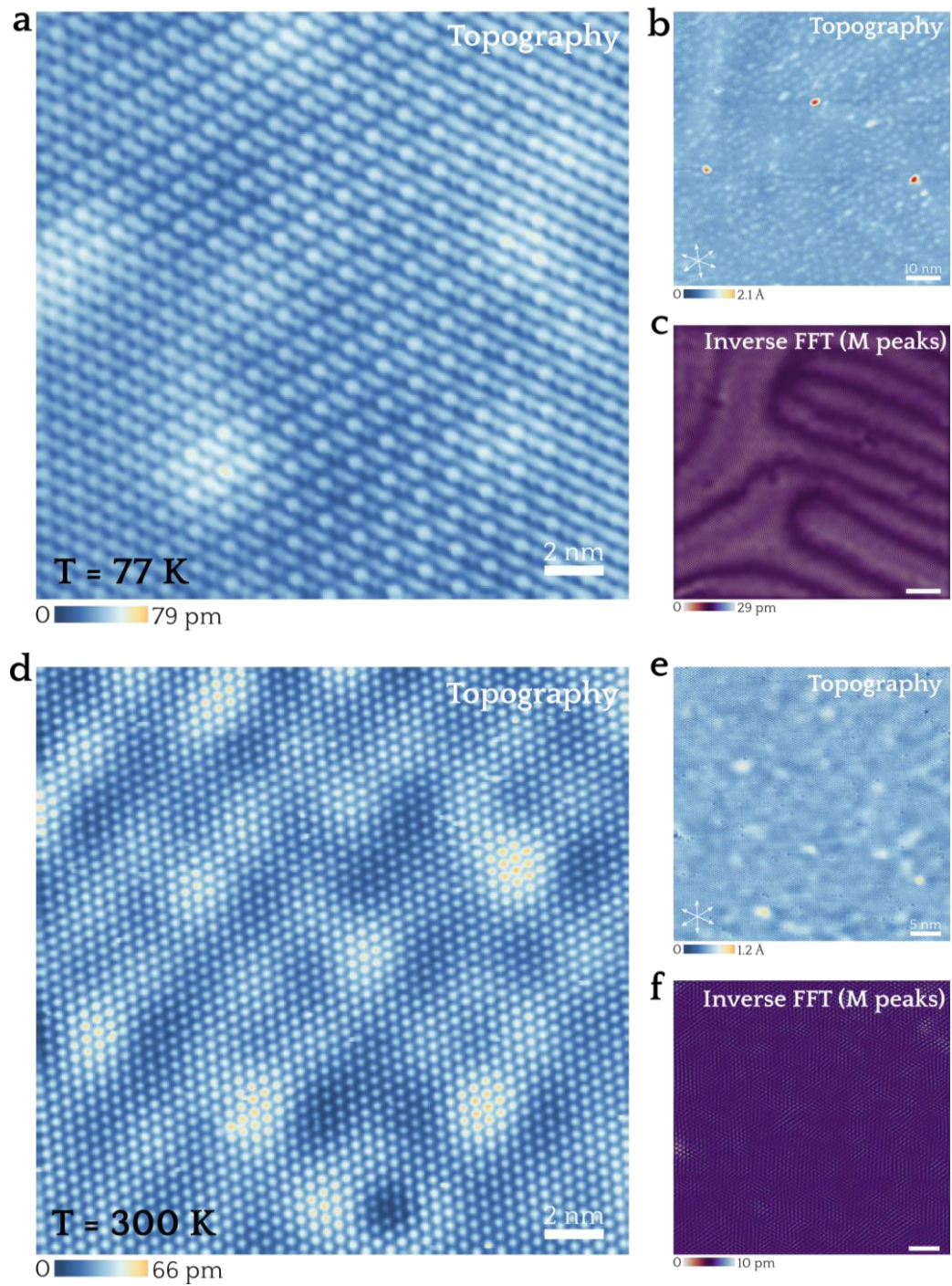

**Fig. S9 | CDW pattern vs. temperature in monolayer TiSe$_2$. a,b**, Short- and long-scale atomically resolved STM images at $T = 77$ K, respectively. Parameters: **a** ($V_s = -0.2$ V, $I_t = 60$ pA), **b** ($V_s = -0.06$ V, $I_t = 100$ pA). **c**, Corresponding I-FFT of the M points. **d,e**, Short- and long-scale atomically resolved STM images at $T = 300$ K, respectively. Parameters: **d** ($V_s = 0.1$ V, $I_t = 600$ pA) and **e** ($V_s = -0.2$ V, $I_t = 1000$ pA). **f**, Corresponding I-FFT of the M points.

## 8. Fourier-filtered STM imaging of relevant periodicities

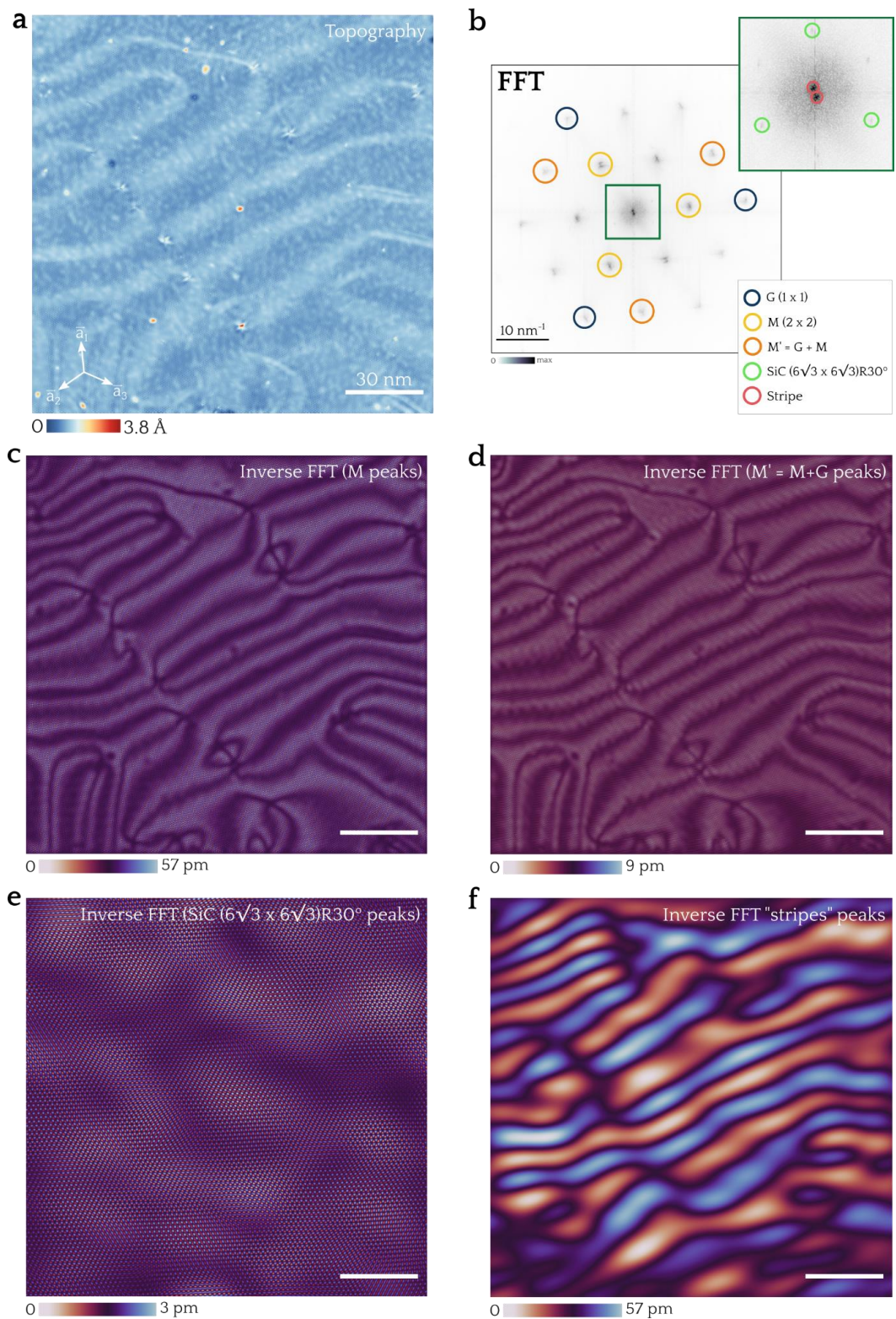

**Fig. S10 | Contribution of other relevant points in *k*-space. a,** Atomically resolved STM image shown in Fig. 2a. **b,** Corresponding FFT with showing the relevant peak features. I-FFT of the M (**c**), M' (**d**), SiC (**e**) and "stripe" (**f**) peaks.

**References**

S1. Chen, P. *et al.* Charge density wave transition in single-layer titanium diselenide. *Nat. Commun.* **6**, 1–5 (2015).

S2. Benesh, G. A., Woolley, A. M. & Umrigar, C. The pressure dependences of $TiS_2$ and $TiSe_2$ band structures. *J. Phys. C Solid State Phys.* **18**, 1595 (1985).